\documentclass[
reprint,
 amsmath,amssymb,
 aps, twocolumn,
 superscriptaddress,prl,citeautoscript,
]{revtex4-1}

\usepackage{graphicx}
\usepackage{dcolumn}
\usepackage{bm}
\usepackage{textcomp}
\usepackage{listings}

\usepackage{color}
\usepackage[usenames,dvipsnames]{xcolor}
\usepackage{hyperref}

\hypersetup{pdftex,linktoc=all,colorlinks=true,bookmarksnumbered}
\hypersetup{citecolor={gray}}
\hypersetup{linkcolor={black}}
\hypersetup{urlcolor={blue}}
\hypersetup{allbordercolors={white}}

\usepackage{gensymb}
\usepackage{lineno}

\newcommand{\unit}[1]{\,\mathrm{#1}}

\newcommand{\etal}{\textit{et al}. }

\makeatletter
\def\fname{\jobname}
\g@addto@macro\fname{.tex}
\makeatother

\immediate\write18{perl ./texcount/texcount.pl -q -sub=section -out=temp.dat \fname}

\begin{document}

\preprint{APS/123-QED}


\title{Guiding of relativistic electron beams in dense matter by longitudinally imposed strong magnetic fields}


\author{M.$\,$Bailly-Grandvaux}
\affiliation{Univ. Bordeaux, CNRS, CEA, CELIA (Centre Lasers Intenses et Applications), UMR 5107, F-33405 Talence, France}

\author{J.J.$\,$Santos}
\email{joao.santos@u-bordeaux.fr}
\affiliation{Univ. Bordeaux, CNRS, CEA, CELIA (Centre Lasers Intenses et Applications), UMR 5107, F-33405 Talence, France}

\author{C.$\,$Bellei}
\affiliation{Univ. Bordeaux, CNRS, CEA, CELIA (Centre Lasers Intenses et Applications), UMR 5107, F-33405 Talence, France}

\author{P.$\,$Forestier-Colleoni}
\affiliation{Univ. Bordeaux, CNRS, CEA, CELIA (Centre Lasers Intenses et Applications), UMR 5107, F-33405 Talence, France}

\author{S.$\,$Fujioka}
\affiliation{Institute of Laser Engineering, Osaka University, 2-6 Yamada-oka, Suita, Osaka, 565-0871, Japan}

\author{L.$\,$Giuffrida}
\affiliation{Univ. Bordeaux, CNRS, CEA, CELIA (Centre Lasers Intenses et Applications), UMR 5107, F-33405 Talence, France}

\author{J.J.$\,$Honrubia}
\affiliation{ETSI Aeron\'autica y del Espacio, Universidad Polit\'ecnica de Madrid, Madrid, Spain}

\author{D.$\,$Batani}
\affiliation{Univ. Bordeaux, CNRS, CEA, CELIA (Centre Lasers Intenses et Applications), UMR 5107, F-33405 Talence, France}

\author{R.$\,$Bouillaud}
\affiliation{Univ. Bordeaux, CNRS, CEA, CELIA (Centre Lasers Intenses et Applications), UMR 5107, F-33405 Talence, France}

\author{M.$\,$Chevrot}
\affiliation{LULI-CNRS, \'Ecole Polytechnique, CEA: Universit\'e Paris-Saclay; UPMC Universit\'e Paris 06:
Sorbonne Universit\'es, F-91128 Palaiseau cedex, France}

\author{J.E.$\,$Cross}
\affiliation{Department of Physics, University of Oxford, Parks Road, Oxford OX1 3PU, UK}

\author{R.$\,$Crowston}
\affiliation{Department of Physics, Heslington, University of York, YO10 5DD, UK}

\author{S.$\,$Dorard}
\affiliation{LULI-CNRS, \'Ecole Polytechnique, CEA: Universit\'e Paris-Saclay; UPMC Universit\'e Paris 06:
Sorbonne Universit\'es, F-91128 Palaiseau cedex, France}

\author{J.-L.$\,$Dubois}
\affiliation{Univ. Bordeaux, CNRS, CEA, CELIA (Centre Lasers Intenses et Applications), UMR 5107, F-33405 Talence, France}

\author{M.$\,$Ehret}
\affiliation{Univ. Bordeaux, CNRS, CEA, CELIA (Centre Lasers Intenses et Applications), UMR 5107, F-33405 Talence, France}
\affiliation{Institut f\"ur Kernphysik, Tech. Univ. Darmstadt, Germany}

\author{G.$\,$Gregori}
\affiliation{Department of Physics, University of Oxford, Parks Road, Oxford OX1 3PU, UK}

\author{S.$\,$Hulin}
\affiliation{Univ. Bordeaux, CNRS, CEA, CELIA (Centre Lasers Intenses et Applications), UMR 5107, F-33405 Talence, France}

\author{S.$\,$Kojima}
\affiliation{Institute of Laser Engineering, Osaka University, 2-6 Yamada-oka, Suita, Osaka, 565-0871, Japan}

\author{E.$\,$Loyez}
\affiliation{LULI-CNRS, \'Ecole Polytechnique, CEA: Universit\'e Paris-Saclay; UPMC Universit\'e Paris 06:
Sorbonne Universit\'es, F-91128 Palaiseau cedex, France}

\author{J.-R.$\,$Marqu\`es}
\affiliation{LULI-CNRS, \'Ecole Polytechnique, CEA: Universit\'e Paris-Saclay; UPMC Universit\'e Paris 06:
Sorbonne Universit\'es, F-91128 Palaiseau cedex, France}

\author{A.$\,$Morace}
\affiliation{Institute of Laser Engineering, Osaka University, 2-6 Yamada-oka, Suita, Osaka, 565-0871, Japan}

\author{Ph.$\,$Nicola\"{i}}
\affiliation{Univ. Bordeaux, CNRS, CEA, CELIA (Centre Lasers Intenses et Applications), UMR 5107, F-33405 Talence, France}

\author{M.$\,$Roth}
\affiliation{Institut f\"ur Kernphysik, Tech. Univ. Darmstadt, Germany}

\author{S.$\,$Sakata}
\affiliation{Institute of Laser Engineering, Osaka University, 2-6 Yamada-oka, Suita, Osaka, 565-0871, Japan}

\author{G.$\,$Schaumann}
\affiliation{Institut f\"ur Kernphysik, Tech. Univ. Darmstadt, Germany}

\author{F.$\,$Serres}
\affiliation{LULI-CNRS, \'Ecole Polytechnique, CEA: Universit\'e Paris-Saclay; UPMC Universit\'e Paris 06:
Sorbonne Universit\'es, F-91128 Palaiseau cedex, France}

\author{J.$\,$Servel}
\affiliation{Univ. Bordeaux, CNRS, CEA, CELIA (Centre Lasers Intenses et Applications), UMR 5107, F-33405 Talence, France}

\author{V.T.$\,$Tikhonchuk}
\affiliation{Univ. Bordeaux, CNRS, CEA, CELIA (Centre Lasers Intenses et Applications), UMR 5107, F-33405 Talence, France}

\author{N.$\,$Woolsey}
\affiliation{Department of Physics, Heslington, University of York, YO10 5DD, UK}

\author{Z.$\,$Zhang}
\affiliation{Institute of Laser Engineering, Osaka University, 2-6 Yamada-oka, Suita, Osaka, 565-0871, Japan}

\date{\today}


\begin{abstract} 

High-energy-density flows through dense matter are needed for effective progress in the production of laser-driven intense sources of energetic particles and radiation, in driving matter to extreme temperatures creating state regimes relevant for planetary or stellar science as yet inaccessible at the laboratory scale, or in achieving high-gain laser-driven thermonuclear fusion. When interacting at the surface of dense (opaque) targets, intense lasers accelerate relativistic electron beams which transport a significant fraction of the laser energy into the target depth. However, the overall laser-to-target coupling efficiency is impaired by the large divergence of the electron beam, intrinsic to the laser-plasma interaction. By imposing a longitudinal $600 \unit{T}$ laser-driven magnetic-field, our experimental results show guided $\geq 10 \unit{MA}$-current of MeV-electrons in solid matter. Due to the applied magnetic field, the transported energy-density and the peak background electron temperature at the $60 \unit{\micro m}$-thick targets rear surface rise by factors $\approx 5$, resulting from unprecedentedly efficient guiding of relativistic electron currents. 

\end{abstract}

\maketitle


\begin{figure*}
\center
\includegraphics[width=\textwidth]{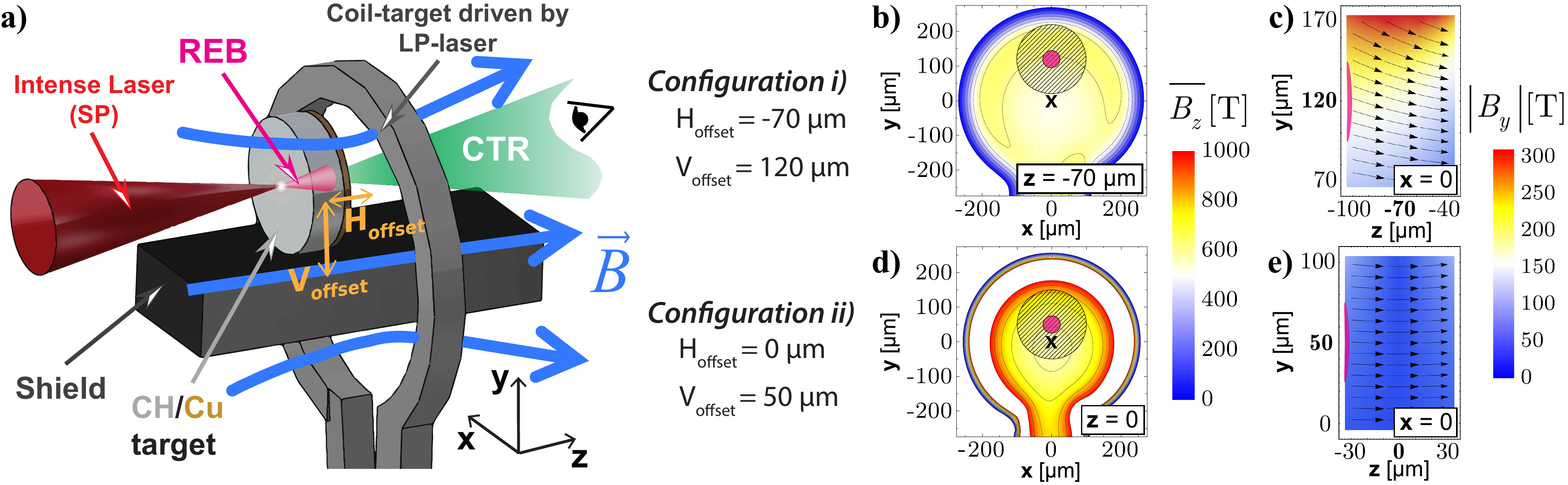}
\caption{\small {\bf Experimental configuration for the study of REB-transport with imposed B-field.} {\bf a)} Sketch of the experimental setup: the REB is generated by the intense SP laser, focused parallel to the coil axis and at normal incidence onto the centre of the front surface of a neighboring solid $50 \unit{\micro m}$-CH / $10 \unit{\micro m}$-Cu thick target of $200 \unit{\micro m}$ diameter. The Ni coil-target (coil radius of $250 \unit{\micro m}$) is previously driven by the ns LP laser. REB-patterns were investigated by imaging the CTR emitted from the transport targets' rear surface. {\bf b)} to {\bf e)} B-field distribution in vacuum at its peak value, $1 \unit{ns}$ after LP-laser driving (origin of the spatial coordinates at the coil center), as experimentally and numerically characterized in \cite{Santos2015}: {\bf b)}, {\bf d)} Amplitude of the B-field longitudinal component averaged over the $60 \unit{\micro m}$ target thickness, $\overline{B_z}$, at the two explored positions of the transport target. The dashed circles represent the position of the transport target in the perpendicular plane: the coil axis and the SP laser axis are respectively represented by the cross-signs and the center of the pink circles, which radius corresponds to the REB initial radius, $r_0$, in the REB-transport simulations. {\bf c)}, {\bf e)} Absolute value of the B-field vertical component $\lvert B_y \lvert$ (color scale) and arrow representation of the B-field lines over the target $x=0$ -slice, for the two target positions. The plots correspond also to the B-field embedded into the targets as initial conditions for the REB-transport simulations in magnetized conditions, in agreement with predictions of the B-field resistive diffusion.}
\label{fig:Fig1}
\end{figure*} 

Production of high-energy-density (HED) flows through solid-density or denser matter is a major challenge for improving laser-driven sources of energetic particles and radiation~\cite{Ledingham2010}, or for optimizing the isochoric heating of dense matter of major relevance in high-gain inertial confinement fusion (ICF)~\cite{Robinson2014, Norreys2014}, in the study of stellar opacities~\cite{Hoarty2013} or of warm dense matter states~\cite{Perez2010}. 
When interacting with dense targets, intense laser pulses drive high-current relativistic electron beams (REB), which can transport a significant fraction of the laser energy into the targets's depth~\cite{Wharton1998, Nilson2010, Westover2014}. However, the energy-density flux degrades rapidly against the penetration depth due to resistive and collisional energy losses~\cite{Pisani2000, Tikhonchuk2002, Santos2007, Vaisseau2015} and mostly to the intrinsically large divergence of the REB~\cite{Santos2002, Stephens2004, Green2008}, as a result of the laser-plasma interaction and the development of electromagnetic instabilities at the target surface~\cite{Adam2006, Debayle2010}. Devising means of controlling the REB transverse spread and confine its propagation within a small radius would be greatly beneficial in the aforementioned research fields and applications. For example, in the framework of the Fast Ignition (FI) scheme for ICF~\cite{Tabak1994, Atzeni2005}, imposed axial magnetic-fields (B-fields) in the $1-10 \unit{kT}$ range should be able to guide $\mathrm{GA}$ currents of $\mathrm{MeV}$ electrons over $100 \unit{\micro m}$ distances from the laser-absorption region, i.e., up to the dense core of nuclear fuel~\cite{Strozzi2012, Wang2015}. This would enhance the electrons kinetic energy coupling to the core and could significantly reduce the ignitor-laser energy needed to initiate nuclear-fusion reactions, potentially leading to high-gain fusion energy release.

Radially-confined REB transport has been experimentally reported due to self-generated resistive B-fields by using specific laser irradiation schemes and/or target structures~\cite{Kar2009, Ramakrishna2010, Perez2011, Scott2012}. The common principle is that collimating B-fields are induced by REB intense currents in resistive media, due either by the beam inhomogeneity or by radially-converging gradients of resistivity along the REB propagation axis~\cite{Bell2003, Robinson2007}. Nonetheless, many electrons are not magnetically-trapped, maintain their initial divergence and are scattered through collisions. In both cases, the number of guided electrons remains under $50\%$. Moreover, proposed improvements involve sophisticated target structures \cite{Robinson2012, Schmitz2012, Debayle2013}, which efficiency is unproven in the harsh conditions of an ICF target. 

Here we present an alternative and apply for the first time an external B-field enough strong to guide MeV-electrons aligned with the REB propagation axis in solid targets. The $600\,$T B-field was produced by an all-optical technique using laser-driven coil targets~\cite{Daido1986, Courtois2005, Fujioka2013, Santos2015, Law2016} (see Methods). This technique creates a stable magnetic field that is of sufficiently long duration to fully magnetize the transport target prior to REB generation. Our results clearly show efficient REB-guiding and increase in energy-density-flux at the rear surface of solid-density targets of $60 \unit{\micro m}$ thickness.



The experiments were conducted at the LULI pico 2000 laser facility with a $1.06 \unit{\micro m}$ wavelength ($1\omega_0$) dual laser beam configuration: i) a high-energy long-pulse beam [LP: $1 \unit{ns}$, $500\pm30 \unit{J}$, $(1.4\pm0.6)\times10^{17} \unit{W/cm^2}$] focused into Ni coil-targets produced a B-field of several hundreds Tesla and duration of a few ns~\cite{Santos2015}, ii) at different delays $\Delta t$ with respect to the LP, a high-intensity short-pulse beam [SP: $1\,$ps, $46 \pm 5 \unit{J}$, $(1.5$ - $3)\times10^{19} \unit{W/cm^2}$] focused at normal incidence generated a REB in solid plastic targets, located at the coil vicinity. The setup at the coil vicinity is sketched in Fig.\,\ref{fig:Fig1}-a). Further details on the coil-targets geometry and laser irradiation are given in Methods.

The REB transport targets were $200 \unit{\micro m}$-diameter and $50 \unit{\micro m}$-thick plastic (CH) cylinders with a $10 \unit{\micro m}$-thick Cu-coating on the rear side. The cylinder's axis was invariably parallel to the coil axis, and for the two experimental runs, we explored successively positioning the target i) shifted from the coil plane (with horizontal and vertical offsets of the target centre with respect to the coil centre of $H_{\mathrm{offset}}=-70 \unit{\micro m}$, $V_{\mathrm{offset}}=120 \unit{\micro m}$), and ii) at the coil plane ($H_{\mathrm{offset}}=0 \unit{\micro m}$, $V_{\mathrm{offset}}=50 \unit{\micro m}$). This enabled us to explore two different 3D spatial distributions for the B-field imposed to the transport targets, as seen in Fig.\,\ref{fig:Fig1}-b) and c) for configuration i) and Fig.\,\ref{fig:Fig1}-d) and e) for configuration ii). For each of the two configurations, the choice of $\Delta t$ controlled the time allowed for B-field diffusion in the transport targets prior to REB injection, testing REB-transport in different conditions of target magnetization. 

The evolution of the transport-target magnetization has been predicted by simulations of the B-field resistive diffusion inside the target as the B-field rises up to its peak-value (rise-time of $\approx 1 \unit{ns}$, consistent with the duration of the LP-laser driver). The results show that by $\approx 1 \unit{ns}$ the transport targets are fully magnetized: the B-field spatial distribution inside the target is similar to the distribution expected in the vacuum case (see Methods).
This magnetization time agrees with a simple linear estimation of the B-field diffusion time $\tau_{\mathrm{diff}} = \mu_0 L^2 / \eta \approx 1 \unit{ns}$ over the length $L=50 \unit{\micro m}$ of the target CH-layer assuming a constant resistivity of $\eta=10^{-6} \unit{\ohm m}$ (expected for CH at $1 \unit{eV}$). 


\begin{figure*}
\center
\includegraphics[width=\textwidth]{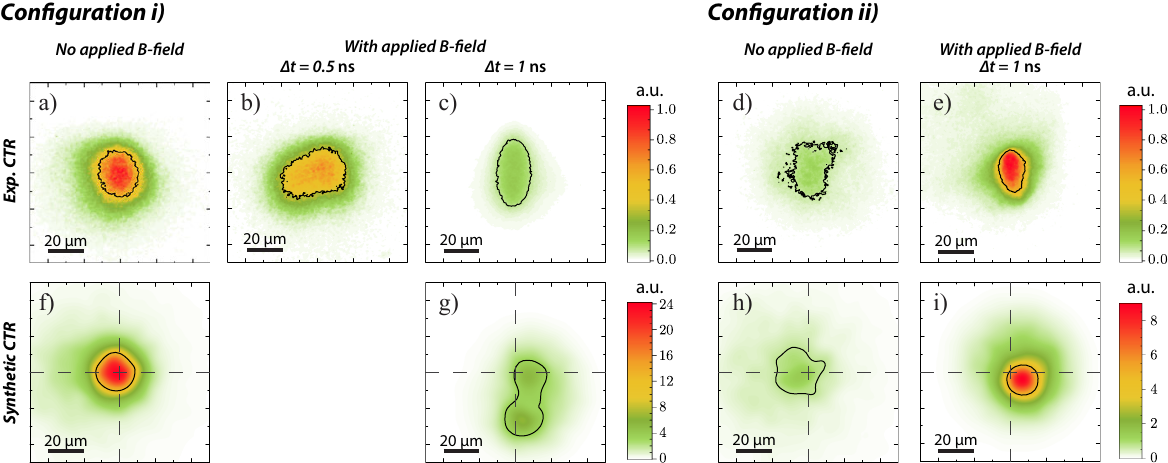}
\caption{\small {\bf Results of the experimental CTR imaging (first row) and of the synthetic CTR calculated from 3D-PIC hybrid simulations of fast electron transport (second row)}, for the two configurations i) target out of the coil plane and ii) target at the coil plane (see Fig.\,\ref{fig:Fig1}), with and without imposed B-field. The contour lines correspond to the half-height of the signals. The crossed dashed lines indicate the position of REB-injection at the targets' front surface.}
\label{fig:Fig2}
\end{figure*}

The REB transverse pattern after crossing the target thickness was investigated by imaging the Coherent Transition Radiation (CTR) emission from the rear surface at twice the SP-laser frequency, $2\omega_0$. The emitting surface was imaged at a $22.5^\circ$ horizontal angle from the target normal into an optical streak camera used with a wide slit aperture as a fast gated 2D frame grabber~\cite{Santos2002}.
Sample results of the CTR signals are shown in the first row of Fig.\,\ref{fig:Fig2}, for target position configurations i) on the left, and ii) on the right, with and without imposing an external B-field, as labelled. The aspect ratio of the signals has been corrected from the observation angle.
For the two data sets, the average SP laser energy and intensity were respectively i) $47\pm 6\,$J and $3.0\pm0.8 \times10^{19}\unit{W/cm^2}$, ii) $49\pm 1\,$J and $1.5\pm 0.4 \times10^{19}\unit{W/cm^2}$. The difference in laser intensity is mainly due to different focal spots in the two independent experimental runs.

Without externally imposed B-field [Fig.\,\ref{fig:Fig2}-a) and d)], we obtained rather large ($\approx 14 \pm 2 \unit{\micro m}$ half-width-half-maximum, HWHM) and fairly-symmetric CTR patterns. When imposing the longitudinal B-field for the target position i) and varying the delay of REB-injection, at $\Delta t=0.5 \unit{ns}$ [Fig.\,\ref{fig:Fig2}-b)] the CTR yield is slightly weaker and its pattern looks twisted yet the average size is comparable to the case without B-field. As mentioned before, the target should not be yet fully magnetized.

At $\Delta t=1 \unit{ns}$ we have obtained CTR patterns significantly different than the case without B-field, for both target positions i) and ii) [respectively Fig.\,\ref{fig:Fig2}-c) and e)]. The CTR-patterns are clearly narrower horizontally. Vertically, the signal is also narrower for configuration ii), while it is elongated for configuration i). These correspond to half-height areas of equivalent radius $\approx 13 \unit{\micro m}$ for configuration i) and $\approx 9 \unit{\micro m}$ for configuration ii). The CTR yield decreased for configuration i) and increased for configuration ii) relative to the corresponding signals without B-field. As discussed above, the delay $\Delta t=1 \unit{ns}$ corresponds to REB transport in magnetized targets. The differences in the patterns shape and yield between Fig.\,\ref{fig:Fig2}-c) and e) are then related to the different B-field distributions inside the targets, as represented in Fig.\,\ref{fig:Fig1}-c) and e): beam electrons are {\it trapped} and follow B-field lines if their Larmor radius becomes comparable or smaller than the beam radius. A $600\,$T field [see values of $\overline{B_z}$ around the REB axis in Fig.\,\ref{fig:Fig1}-b) and d)] is strong enough to {\it confine} MeV-electrons. It is worth noting that the $B_y$-component in config. i) [Fig.\,\ref{fig:Fig1}-c)] deviates downwards the REB propagation axis. 
The consequent inclined REB path in the transport target results in an observed yield drop on the detected CTR. This behaviour of the CTR is reproduced and will be further explained in the following part of the article, by means of REB transport simulations.
In comparison, for config. ii) with optimum imposed B-field symmetry, the narrowing of the signal e) compared to d) is significant and is a signature of a radially-pinched REB. Here the REB axis is not deviated. As a result, an increase in the REB density occurs and is directly observed as an enhancement of the CTR yield by a factor 6. 


\begin{figure*}
\center
\includegraphics[width=\textwidth]{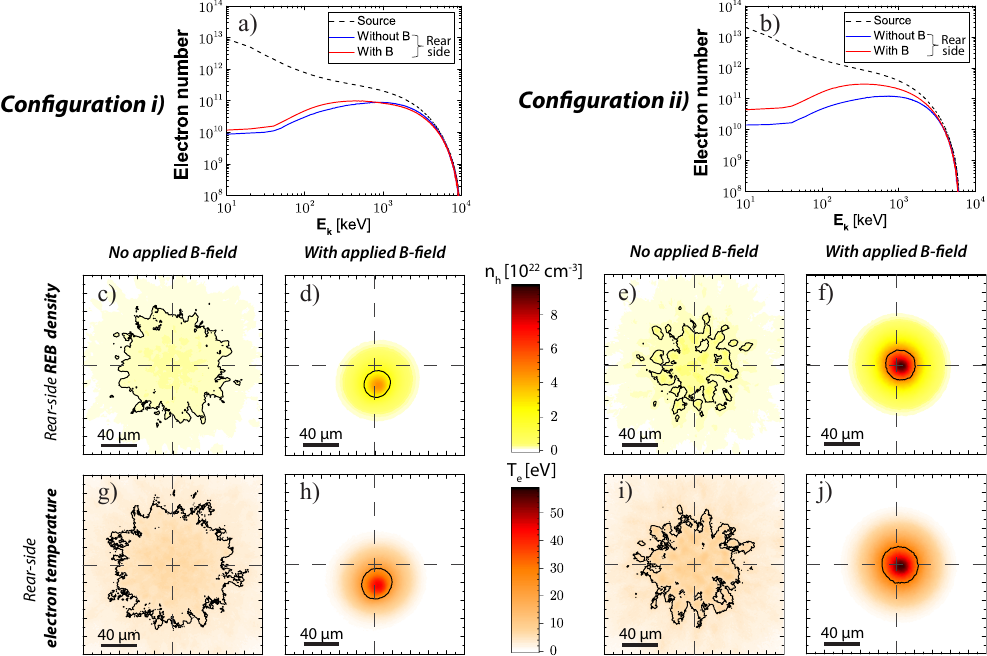}
\caption{\small {\bf REB energy spectra (first row), rear-side REB density (second row) and background electron-temperature (third row) unfolded from 3D PIC-hybrid transport simulations.} The results are plotted for target-position configurations i) on the left and ii) on the right, without and with B-field. {\bf a)}, {\bf b)} REB energy spectra at the targets' front-side (source, dashed lines) and rear-side (full lines). {\bf c)} - {\bf f)} REB density at the target rear surface (time-integrated). {\bf g)} - {\bf j)} Final background electron-temperature at the target rear surface.
In {\bf c)} - {\bf j)} the contour lines correspond to the half-height of the signals and the crossed dashed lines indicate the position of REB-injection at the targets' front surface.}
\label{fig:Fig3}
\end{figure*}

Details of the REB propagation and energy transport were unfolded by simulations using a 3D PIC-hybrid code for the electron transport~\cite{Honrubia2005}, accounting for fast electron collisions with the background material and REB self-generated fields. We simulated REB transport both with and without imposed B-fields. For the former case, we assumed full target magnetization (corresponding to the experimental $\Delta t = 1\,$ns) imposing as initial conditions the 3D B-field distributions illustrated in Fig.\,\ref{fig:Fig1}-b)-e).

The initial REB total kinetic energy was set to $30\%$ of the on target SP-laser energy, and injected at the front surface over a region of $r_0\approx 25 \unit{\micro m}$-radius HWHM, corresponding to empiric factors 4 or 3 of the SP-laser focal spot size HWHM, respectively for config. i) or ii). The injected electron kinetic energy spectra [dashed lines in Fig.\,\ref{fig:Fig3}-a) and b)] were characterized by power laws for the low energy part $\propto (E_k)^{-1.6}$ and exponential laws for the high energy part $\propto \exp{\left( -{E_k}/{T_h} \right)}$ with $T_h^\mathrm{i)}=2.0\,$MeV and $T_h^\mathrm{ii)}=1.3\,$MeV, as predicted by the ponderomotive potential for the corresponding laser parameters~\cite{Wilks1992}. The injection angular distribution was characterized by a $30 \degree$ mean radial angle and a $55 \degree$ dispersion angle as defined in~\cite{Debayle2013, Vauzour2014} (see Methods). All the above geometric and energy REB source parameters are consistent with our previous characterization in the same laser facility and equivalent laser parameters~\cite{Vauzour2012,Vauzour2014}.

For a direct comparison with the experimental data, we developed a synthetic CTR-emission post-processor applied to the transport code output. CTR is reconstructed by the coherently added transition radiation fields produced by each simulated macro-particle. The yield, spatial and angular distributions of CTR are non-linearly dependent on the density of fast electrons crossing the target-vacuum boundary, on their momentum distribution (norm and angle relative to the target normal) and on the observation angle relative to the symmetry axis of the momentum distribution~\cite{Bellei2012}, as illustrated in the Methods. 
Further details on the parameters and assumptions of REB-transport simulations and CTR post-processing are found in the Methods.

Synthetic CTR-signals are presented in Fig.\,\ref{fig:Fig2}-f), g), h) and i), reproducing fairly well the experimental CTR patterns as well as the relative signal yield change when imposing the B-field. In more details, the simulations reproduce with ${15 \pm 2\%}$ relative errors the ratio of CTR-yield (with B-field / without B-field) for both target positions. By applying the B-field, the synthetic CTR emission is pinched when the target is placed at the coil plane [Fig.\,\ref{fig:Fig2} i)] and the signal is vertically elongated when the target is placed outside the coil plane [Fig.\,\ref{fig:Fig2} g)]. The experimental CTR patterns' radius (azimuthally averaged) are reproduced with $15 \pm 5\%$ relative errors for config. i) and ii), except when the target is placed outside the coil plane with applied B-field: the elliptic shape of the experimental signal is not exactly reproduced and its averaged radius is reproduced with a $\approx 33 \unit{\%}$ relative error.

As for the REB-transport, Figure\,\ref{fig:Fig3} shows the corresponding simulation results at the $60 \unit{\micro m}$-thick targets' rear side surface. Figure\,\ref{fig:Fig3} a) and b) show the time-integrated REB kinetic energy spectra $E_k$ for simulations with (full red) and without (full blue) B-field, compared to the spectrum at the front surface (dashed black). Beam electrons with energy $E_k<100\,$keV are absorbed or scattered out of the simulation box before crossing the full target thickness as expected from the direct collisions with the background material and by resistive field effect linked to the neutralizing return current of thermal electrons~\cite{Vauzour2012, Vauzour2014}. In config. ii) the symmetric B-field slightly mitigates energy losses for $E_k$ up to $\approx 1\,$MeV. 

The time-integrated REB-density patterns in [Figure\,\ref{fig:Fig3} c)-f)] clearly show radial pinching due to the imposed B-fields, increasing the peak density by factors $\approx 15$ and $\approx 20$ and decreasing the beam mean radius by factor $\approx 3$ and $\approx 2$, respectively for configurations i) and ii). An other encouraging outcome is that the imposed B-field smoothes the REB-filaments compared to simulations without B-field, for both target positions. Yet, it is noticeable in simulations that the B-field asymmetry in configuration i) deviates vertically the REB from its injection axis, exiting the target with a vertical shift of $\approx 23\unit{\micro m}$ with respect to its injection axis and an angle of $\approx 20^\circ$. On one hand, this vertical deviation broads both time-of-flight and momentum-angle of the beam electrons at the target rear side. This contributes to the coherence loss of the CTR [Fig.\,\ref{fig:Fig2}-c), g)] compared to the unmagnetized transport case [Fig.\,\ref{fig:Fig2}-a), f)]. On the other hand, the deviation points the beam away from the CTR collecting lens. Calculations of the CTR yield variation as a function of the collecting lens position are given in Methods. The chosen lens position in our setup accounts for half of the total CTR signal drop observed in config i). The remaining part of the CTR drop is due to the additional broadening of the electron bunches and consequent CTR loss of coherence.

The substantially smoother, narrower and denser beams in Fig.\,\ref{fig:Fig3}-d) and f) correspond to unprecedented efficient guiding and improved energy-density flux. The impact is clearly seen in the reached peak background electron temperature [Figure\,\ref{fig:Fig3} g) - j)], higher with B-field by a factor $\approx 5.9$ for both target positions. 

\begin{figure}
\center
\includegraphics[width=0.99\columnwidth]{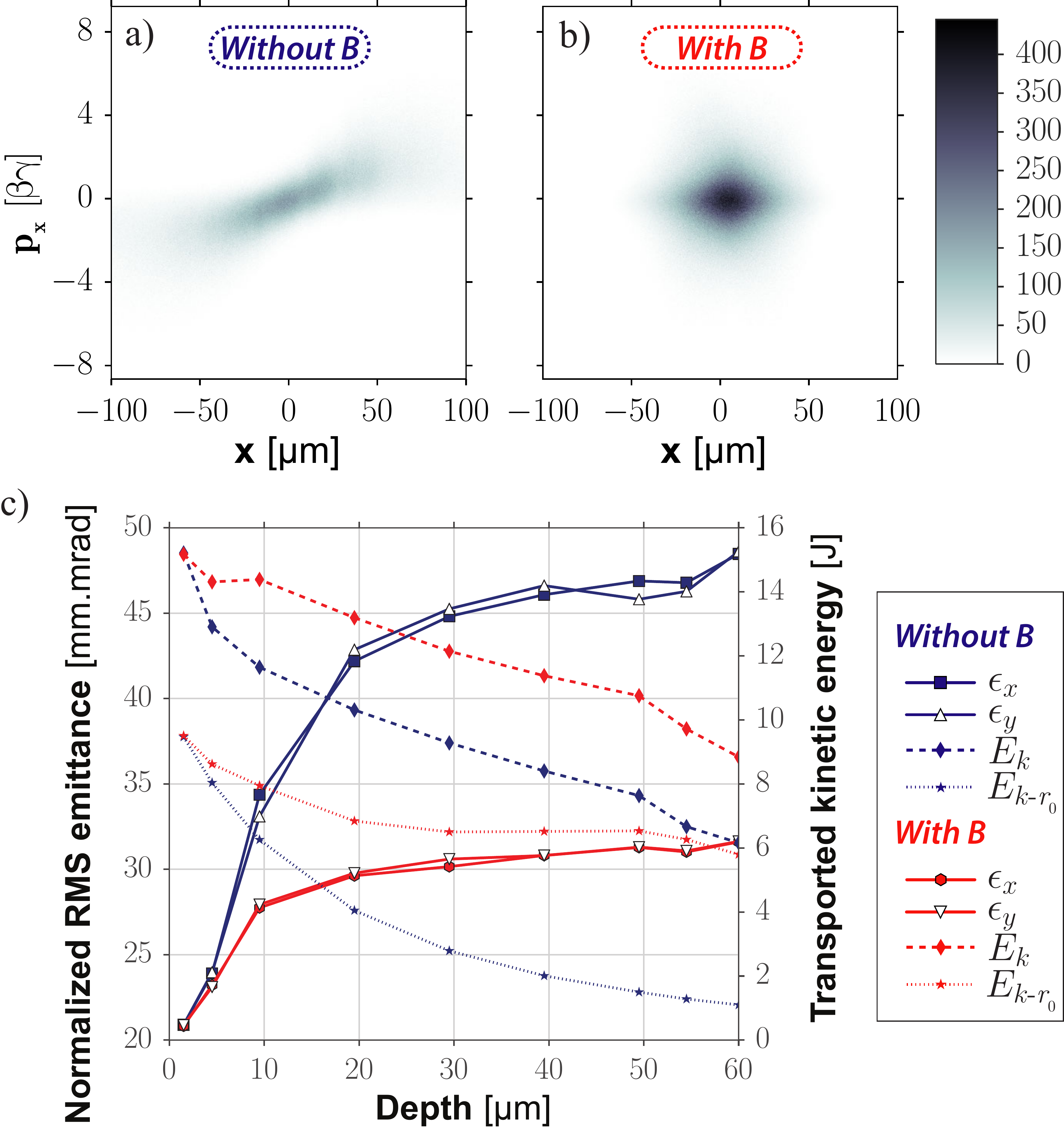}
\caption{\small {\bf REB transverse phase-space ($x$, $p_x$) at the target's rear surface and REB emittance and total kinetic energy as a function of its propagation depth, for configuration ii) with transport targets at the coil plane: cases with (red) and without (blue) imposed B-field}. a) REB phase-space ($x$, $p_x$) at the target rear-surface, for the case without B-field. b) Idem for the case with B-field. c) REB normalized RMS emittance in horizontal $\epsilon_x$, and vertical $\epsilon_y$, transverse directions (left-hand side ordinates); $E_k$ and $E_{k-r_0}$ respectively the total transported kinetic energy (diamonds) and its fraction within the initial REB radius $r_0$ centred in target axis (stars), as a function of the propagation depth (right-hand side ordinates).} 
\label{fig:Fig4}
\end{figure}

Guiding of the REB by an imposed B-field is further evidenced by analysing the phase-space maps of the electrons reaching the targets' rear surface. For configuration ii), the transverse horizontal coordinates phase-space ($x$, $p_x$) is plotted in Fig.\,\ref{fig:Fig4}: in a) without B-field, the inclined shape of high ellipticity is characteristic of symmetric correlated transverse momentum and position of a regularly diverging beam; in b) with B-field, the phase-space map is significantly narrower than in a) but only for the spatial-coordinates. The electrons cyclotron movement de-correlates positions and momenta and their radial spread is limited as they are trapped and flow along (rotating around) the B-field lines, as expected from the guiding criterium where the $\approx 10\unit{\micro m}$-Larmor radius, calculated for the REB-source mean kinetic energy $\approx 1\,$MeV and a $600\,$T B-field, is smaller than the REB-source radius $r_0\approx 25\unit{\micro m}$. Yet the B-field does not really affect the electrons divergence as the width of the transverse momenta distribution is maintained. 

Figure~\ref{fig:Fig4}-c) summarizes the evolution with target depth of the REB normalized root-mean-square (RMS) emittance in both horizontal ($\epsilon_x$) and vertical ($\epsilon_y$) transverse coordinates (left-hand side ordinates; for further details on $\epsilon$ calculation, see Methods). The plot clearly shows the effect of the imposed B-field: without B-field the electron beam emittance rises considerably due to collisional scattering (full blue lines), an effect considerably mitigated in case of magnetic guiding of the REB (full red lines).The emittance evolution over the first $\sim 10\unit{\micro m}$ is particularly steep for the unmagnetized case, this is related to the high diffusivity of the lower-energy electrons in the REB. For larger depths, the two slopes are less steep, as the low energy electrons are previously lost by collisional scattering and Ohmic heating~\cite{Vauzour2012, Vauzour2014, Vaisseau2015}.\\ \\
\indent To understand the energy transport, we plot in the same graph (right-hand side ordinates) the evolution of the time-integrated REB total energy ($E_k$, diamonds connected by dashed lines) and REB energy encircled over the surface corresponding to the initial REB-source, $\pi r_0^2$ ($r_0$ is the initial REB radius) kept centred with the injection axis ($E_{k-r_0}$, stars connected by dotted lines). The $E_k$-loss rate against target depth is comparable for the two cases without (blue) and with (red) B-field, except for the first $\sim 10\unit{\micro m}$ where the B-field efficiently confines electrons and also smoothes REB filamentation. About 45\% more energy is transported to the target rear in the magnetized case due to the magnetic confinement mitigating the high diffusivity of low-energy particles.
Much more importantly, as a consequence of REB guiding in the magnetized case, the $r_0$-encircled energy around the injection axis at the target rear contains $\approx 66\%$ of the total transported energy, against only $\approx 18\%$ for the unmagnetized case. As a consequence, the REB energy-density flux after crossing the target thickness increases by $\approx 5.3 \times$ by applying the B-field.


In conclusion, we have succeeded to efficiently guide a laser-accelerated MeV electron beam through solid-density matter by, for the first time, imposing a $600\,$T B-field parallel to the electron beam propagation axis. The B-field was generated by an all-optical process using a coil-target driven by a high-energy ns laser interaction~\cite{Santos2015}. This B-field was driven $1\,$ns before the REB acceleration, providing a sufficient time for the magnetization of the transport target. 
In our best setup configuration, we found at the rear surface of a $60 \unit{\micro m}$ target that the energy density transported by the fast electrons and the peak background electron temperature increase respectively by factors of $\approx 5.3$ and $\approx 5.9$ compared to the case without imposed B-field. This unprecedented enhancement in energy-density transport through dense matter is notable when compared to experiments based on the REB-guiding by self-generated resistive B-fields~\cite{Kar2009, Ramakrishna2010, Perez2011, Scott2012}.  
Our experimental results set the ground for laboratory studies in new regimes of matter opacities and equations of state at extreme temperatures. In the particular context of laser-fusion research, relevant experiments with target compression in magnetized conditions and magnetically-guided relativistic electron beams should potentially optimize energy coupling to the dense core of nuclear fuel~\cite{Robinson2014, Wang2015, Johzaki2015, Honrubia2016, Azechi2016}.


\section*{Methods}

\subsection*{Production of strong magnetic-fields driven by laser}

\begin{figure}
\center
\includegraphics[width=\columnwidth]{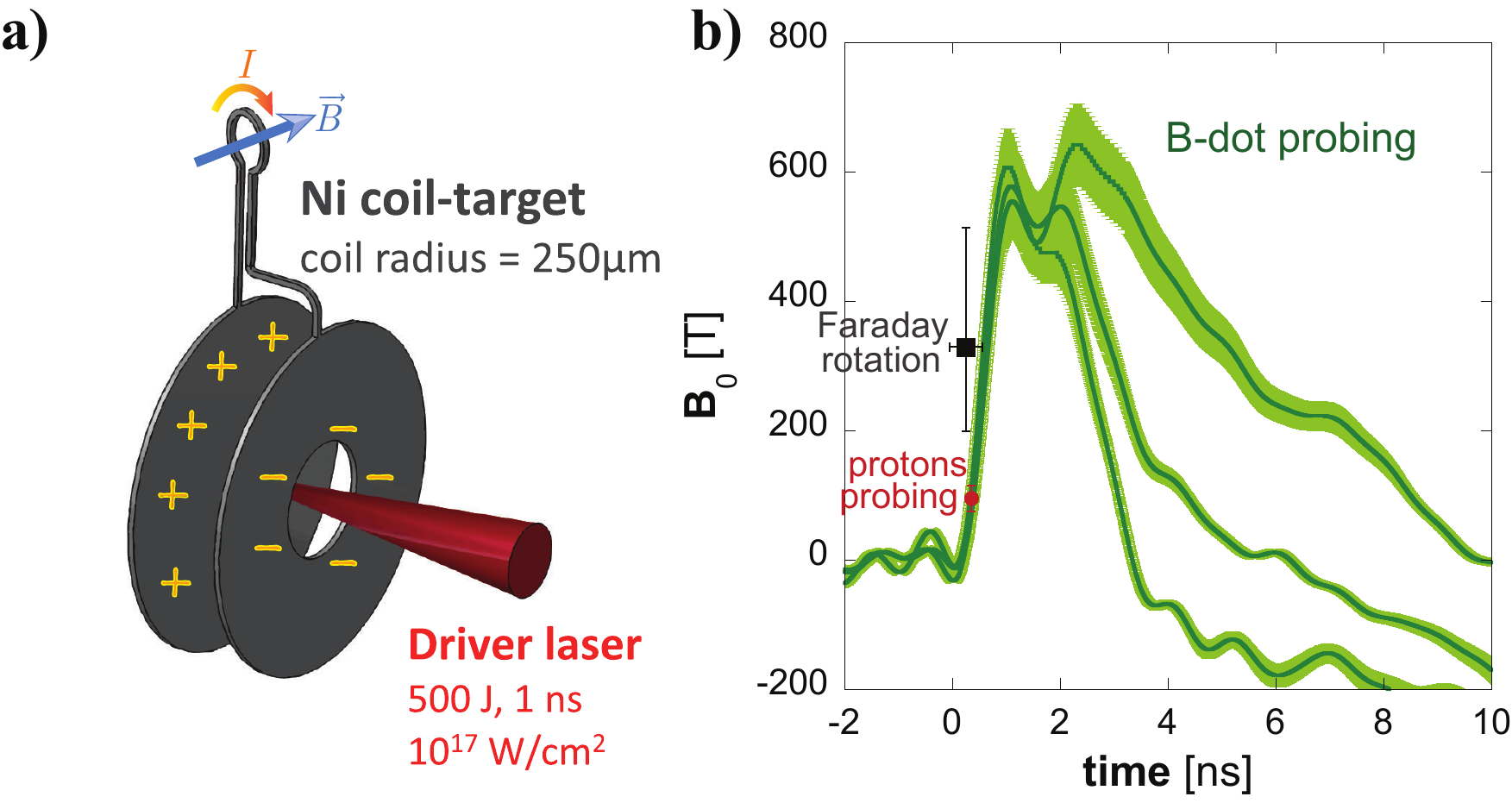}
\caption{\small {\bf Production of strong magnetic-fields driven by laser.} {\bf a)} Illustration of the B-field production mechanism with laser-driven coil-targets. {\bf b)} Experimental results for the B-field strength at the center of the coil as a function of time, inferred from measurements by B-dot probes (green curves), Faraday rotation (black square) and proton-deflectometry (red circle).
\label{fig:Fig5}}
\end{figure}

The first part of our experimental campaign was devoted to producing and characterizing strong B-fields~\cite{Santos2015}. We used a LP laser with $1.06 \unit{\micro m}$ wavelength, $500\pm30 \unit{J}$ energy, $1 \unit{ns}$ flat-top duration ($\approx 100 \unit{ps}$ rise time), focused into loop-shaped targets. These were made out of $50 \unit{\micro m}$-thick Ni-foil, laser-cut and bent to form two parallel disks, connected by a coil-shaped wire of $500 \unit{\micro m}$-diameter [see Fig.\,\ref{fig:Fig5}-a)]. The laser was incident along the surface normal of one disk, passing through the hole of the other, yielding $(1.3\pm0.8)\times10^{17} \unit{W/cm^2}$ of focused intensity. The target is charged by the laser-generated supra-thermal electrons escaping its potential barrier. We believe also that some of them are captured on the opposite holed disk. The resulting discharge current through the wire loop closes the circuit producing a quasi-static (time-scale of a few ns), dipole-like B-field in the coil region. The laser-charging and discharge through the wire process during the laser irradiation, after this the target discharges like an RL-circuit. Figure\,\ref{fig:Fig5}-b) shows results for the B-field strength at the coil center, $B_0$, as a function of time, inferred from induction measurements at $7 \unit{cm}$ from the coil, approximately along the coil axis, using $2.5 \unit{GHz}$ bandwidth pick-up B-dot probes (green curves). The results are related to the discharge current intensity $I$ and to the coil radius $a$, according to $B_0Ê\approxÊ\mu_0I/2a$. $\mu_0$ is the vacuum permeability. The inferred results show reproducible charging-time, consistent with the driver LP laser duration, and peak value of $\approx 600 \unit{T}$ ($\pm 10\%$). For the earlier probing times, the B-field strength was separately measured using the Faraday rotation effect on the polarization of a probe laser beam (black square) and of the deflections of an energetic proton-beam (red circle) produced by the interaction of the SP beam with a Au foil placed a few mm away from the coil. These measurements confirm the inferred results using the B-dot pickup coils, at least for early time. The produced B-fields have a dipole-like spatial distribution over a $1 \unit{mm^3}$-scale volume. The spatial-integrated energy of the B-field at peak-time corresponds to $\approx 4.5\%$ of the driver laser energy. Further details are described in reference~\cite{Santos2015}.

\subsection*{Simulations of the magnetic-field diffusion over the transport targets}

\begin{figure}
\center
\includegraphics[width=\columnwidth]{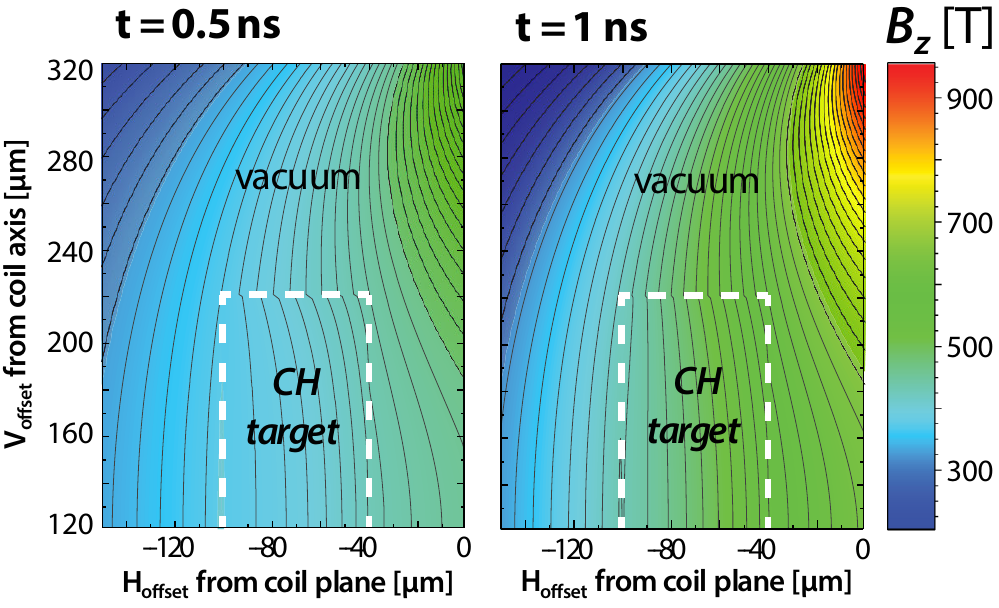}
\caption{\small {\bf Axis-symmetric simulations of the B-field resistive diffusion over the transport targets.} The B-field spatial distribution is given at $\Delta\tau=0.5 \unit{ns}$ (left panel) and at $\Delta\tau=1\unit{ns}$ (right panel) for the target positioning configuration with $H_{\mathrm{offset}}=70\unit{\micro m}$ and $V_{\mathrm{offset}}=120\unit{\micro m}$. From the breaking on the contour lines at the target-vacuum interface, we conclude that there is a discontinuity of the B-field for $\Delta\tau=0.5\unit{ns}$ and that discontinuity practically disappears at $\Delta\tau=1\unit{ns}$, assumed as the time of full target magnetization.
\label{fig:Fig6}}
\end{figure}

The transport-target magnetization has been predicted by simulations of the B-field resistive diffusion inside the target \cite{Honrubia2016}, as the B-field rises with time up to its peak-value. The model describes the penetration of external B-fields in the target material, according to the diffusion equation $\partial_t \vec B=\eta\nabla^2B/\mu_0$, which is valid for small magnetic Reynolds number. At each time iteration, Amp\`ere's law calculates the induced current density, from which ohmic heating is then computed for the temperature map of the target along with a new resistivity map. 
Figure\,\ref{fig:Fig6} presents sample results for the target position configuration i), at $t=0.5\unit{ns}$ (left panel) and $t=1 \unit{ns}$ (right panel). At $t\approx 1 \unit{ns}$ we consider the target to be fully magnetized, as the B-field spatial distribution inside the target becomes comparable to what is simulated in vacuum at the same position. Further, the B-field gradients discontinuities practically disappear at the target edges. 

\subsection*{PIC-hybrid simulations of electron beam transport}

PIC-hybrid simulations allow to describe fast electron beam transport in dense matter, where the injected beam current is modeled kinetically by a particle-in-cell (PIC) method and the neutralizing return current of background thermal electrons is described by Ohm's law as an inertialess fluid \cite{Davies1997, Gremillet2002, Honrubia2005}. The hybrid method neglects high frequency effects, hence enabling a simplification of Maxwell equations by neglecting the Poisson equation and the displacement current in the Maxwell-Ampere equation. 

Our simulation box corresponded to the transport-target dimensions, reproducing its CH-Cu structure in terms of background density and resistivity behaviour as a function of the evolving background electron temperature due to REB-deposited energy. Both collisional and ionization processes were taken into account for computing the evolution of the background resistivity, computed using the classical Drude model according to the Eidmann-Chimier model~\cite{Eidmann2000, Chimier2007} and as described in \cite{Vauzour2014}. The background electron temperature is initiated at $0.1 \unit{eV}$ and $1 \unit{eV}$, respectively for the cases without and with B-field. The higher initial temperature in the later case accounts for the target pre-heating by intense X-rays issuing from the LP-laser interaction and the coil driven by the intense discharge current.  
   
As for the injected REB source parameters, we considered kinetic energy $E_k$ distributions between $8 \unit{keV}$ and $10 \unit{MeV}$, characterized by power laws for the low energy part $\propto (E_k)^{-1.6}$ and exponential laws for the high energy part $\propto \exp{\left( - {E_k}/{T_h} \right)}$ with $T_h^\mathrm{i)}=2.0\,$MeV and $T_h^\mathrm{ii)}=1.3\,$MeV. The characteristic hot-electron temperature is given by the ponderomotive potential~\cite{Wilks1992}, $T_h=m_\mathrm{e}c^2 (\sqrt{1+a_0^2}-1)$, where $a_0$ is the normalized laser intensity. The corresponding fast electron angular distribution was approximated with the following form~\cite{Debayle2013}: $f_h(\theta,r)\propto \exp{\left[ -(\theta-\theta_r)^2/\Delta\theta_0^2\right]}$
with $\Delta\theta_0 = 55 \degree$ the dispersion angle, and $\theta_r=\arctan{(\tan{(\Theta)} \: r/r_0)}$, where $\Theta=30 \degree$ is the mean radial angle at initial REB radius $r_0 \approx 24 \unit{\micro m}$ in configuration i) and $r_0 \approx 27 \unit{\micro m}$ in configuration ii). These geometric and energy REB source parameters are consistent with an experimental characterization made previously in the same facility with equivalent laser parameters and are supported by benchmarked simulations \cite{Vauzour2014}. Approximately 50 millions macro-particles were used to simulate the propagation of $\sim 1-2 \times 10^{14}$ electrons through the target. The total simulation time was set to $3.6 \unit{ps}$ with temporal and spatial resolutions of $3 \unit{fs}$ and $1 \unit{\micro m}$, respectively.

Given the ps-time scale of the REB-transport, very fast if compared to the ns-scale evolution of the B-field strength or of its diffusion in the target, we assumed that the B-field distribution is constant over each simulation run-time. For the cases with applied external B-field, we only considered fully-magnetized targets ($\Delta t=1 \unit{ns}$). The B-field spatial distribution inside the target is calculated as in vacuum by a 3D magnetostatic code~\cite{Radia}, consistently with the experimental characterization of the B-field space-time evolution obtained with laser-driven coil-targets~\cite{Santos2015}. 

\vspace{-1em}
\subsection*{Coherent Transition Radiation as a diagnostic of laser-accelerated relativistic electron beams}

Coherent Transition Radiation (CTR) is produced by the REB crossing the target-vacuum boundary~\cite{Baton2003, Popescu2005}. Its time-scale, of the order of a few ps, follows that of the fast electron flux envelope. Yet, the pulsed character of relativistic laser-acceleration mechanisms modulates longitudinally the REB-current as a comb of periodic micro-bunches. The coherent interference of the transition radiation produced by the electron-comb crossing the rear surface yields peak emissions at the spectral harmonics of the bunch frequency. 

In the present experiment, the CTR imaging system of the transport targets' rear surface was composed by two doublet-lenses with an optical aperture of $f/9$, and was aligned on the equator plane looking at the target rear Cu-surface with a $22.5 \degree$-angle with respect to its normal. The optical system produced images with a magnification of $\approx 20$ with a spatial resolution of $\approx 3\unit{\micro m}-\sigma$. The streak camera was used with an open slit ($\approx 5 \unit{mm}$) and the faster sweep speed of $0.5 \unit{ns}$/screen synchronized to the SP laser beam interaction, allowing to freeze the 2D pattern of the prompt CTR emission from the target surface. $2\omega_0$-light was selected by a $10 \unit{nm}$ FWHM-bandwidth interferometric filter centred at $532 \unit{nm}$. As an extra precaution for reducing the noise level due to any spurious light from the coil-target, the first lens produced an intermediate image where mechanical filtering allowed to select only the central region of the imaged field, corresponding to the REB-transport target surface.

The CTR data was used as benchmarking reference for the electron transport simulations coupled to a synthetic CTR-emission post-processor. As the hybrid transport code continuously injects particles during the laser pulse duration, and as such does not simulate the REB longitudinal (or temporal) modulations, we assumed that the periodic electron micro-bunches produced throughout the duration of the laser-plasma interaction are identical and that the CTR spatial pattern and angular distribution are not dependent on the number of bunches. The CTR is therefore calculated for a single electron bunch, yielding intensities at one given wavelength in arbitrary units. 

In general, the coherence of the radiation can be decomposed into a temporal and a spatial component~\cite{Santos2007, Bellei2012}. For the temporal coherence, the phase difference, $\Phi_i-\Phi_j$ of the transition radiation fields emitted at the target rear side is calculated after considering the time of flight of each individual macro-particle, i.e. we record the time difference between injection and time of arrival at the target rear side. For the spatial coherence, we assume no phase shift between macro-particles belonging to the same cell at the target rear side. This is justified by noting that the cell size in the simulation, of $1\unit{\micro m} \times 1 \unit{\micro m}$, is comparable to the observed wavelength and smaller than the spatial resolution of the experimental imaging system. The finite spatial resolution is taken into account by convoluting the CTR out of the simulation with a $3 \unit{\micro m}$ standard deviation Gaussian function.

The coherent addition of the fields $\vec E_i$ emitted by each macro-particle is therefore given by:

\begin{equation}
I_{TR} \propto \sum_i \lvert \vec E_i \lvert^2 + \sum_i \sum_{j,\, j\neq i} \lvert \vec E_i \lvert \lvert \vec E_j \lvert \exp\left( i(\Phi_i-\Phi_j) \right) \ .
\end{equation}

\begin{figure*}
\center
\includegraphics[width=\textwidth]{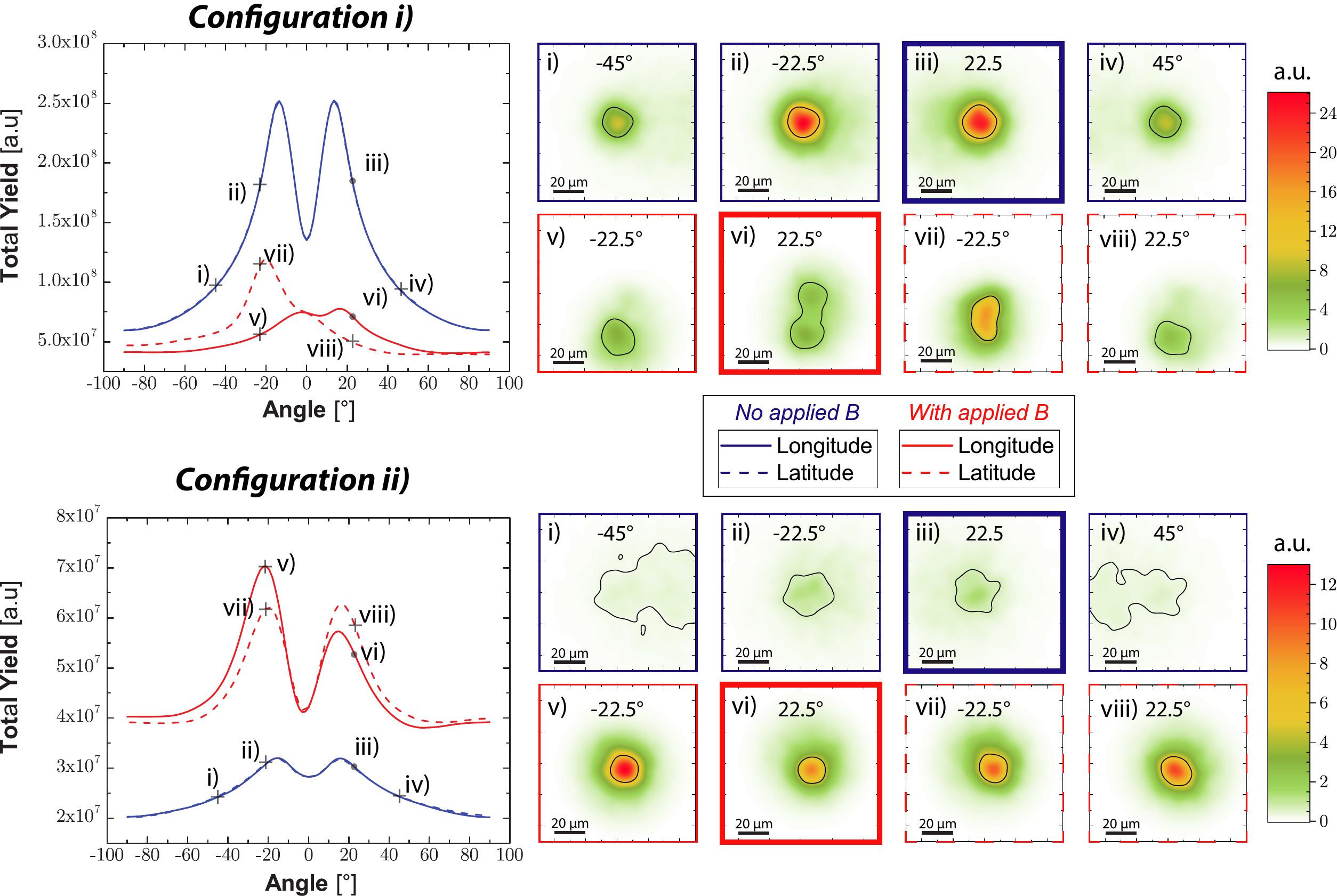}
\caption{\small {\bf Synthetic CTR calculated from the benchmarked simulations of REB transport in the experimental conditions.} The graphs on the left plot the total CTR yield as a function of the lens position, in terms of varying longitude (full lines) or varying latitude (dashed lines). The symbols identify the coordinates corresponding to the sample synthetic CTR images on the right. Blue curves and boxes correspond to REB transport without B-field, red curves and boxes to REB transport with imposed B-field. The four experimental configurations analysed in the article are identified by the thicker frames and by the dot-symbols in the graphs. 
\label{fig:Fig7}}
\end{figure*}

The final synthetic images to be compared to the experimental CTR data take into account the imaging system solid angle, angle of observation and magnification factor. Figure \ref{fig:Fig7} shows results for the synthetic CTR for the four experimental situations analysed in the article: with and without B-field, with in config. i) the transport target outside the coil plane and in config. ii) the transport target at the coil plane. The time and space integrated CTR yields are plotted as a function of the imaging diagnostic orientation with respect to the target normal, by varying both latitude (dashed lines) and longitude (full lines). The figure also shows sample synthetic CTR images, corresponding to the lens coordinates indicated in the graphs. The four images corresponding to the experimental setup (longitude $22.5^\circ$, latitude $0^\circ$) are identified by the thicker box frames and by the dots on the graphs. The size of the dots are representative of the solid angle of the collecting lens in the experiment. The CTR yield calculation obtained either by considering the full solid angle of the lens or its center point differs by less than $1\%$, justifying the point-like lens approximation made for the calculations of Figure \ref{fig:Fig7}.

In the case without applied B-field (blue curves/frames), the diverging REB propagation is symmetric around the injection axis. The result is a symmetric CTR emission pattern. As expected \cite{Bellei2012}, peak emission in the CTR image occurs at an angle close to $1/\gamma$, where $\gamma$ is the relativistic factor corresponding to the mean energy of the REB. With B-field (red curves/frames), the REB propagation is confined to a small radius and revolves at the cyclotron frequency around its symmetry axis. In configuration i), the REB axis deviates downwards due to the inclination in B-field lines, explaining the asymmetric angular dependence of the CTR emission. The peak emission in latitude corresponds to the direction angle of the REB exiting at the target rear. In configuration ii) the angular distribution is closer to that of a beam symmetric to the target normal, yet due to the cyclotron movement, the exiting position and angle are related to the target thickness.

\vspace{-1em}
\subsection*{Calculation of REB-emittance}

Emittance is a quantity of area or volume in phase space of particles. It is usually used as a property to describe a beam along its propagation when the motion of particles in transverse and longitudinal planes are weakly coupled. As in conventional accelerators, fast electron beams accelerated by high intensity lasers have momentum mostly directed in the longitudinal direction. 
On the canonical phase-space, a statistical definition of the normalized Root Mean Square (RMS) emittance in the $(x,p_x)$ plane is given by :
\[{\varepsilon _{x,n,RMS}} = \sqrt {\left\langle {{x^2}} \right\rangle \left\langle {{p_x}^2} \right\rangle  - {{\left\langle {x\,{p_x}} \right\rangle }^2}} \]
It is convenient to use momenta in dimensionless units of $\beta \gamma$ (relativistic parameters).
${\left\langle x^2 \right\rangle }$ defines the second central moment of the particle distribution $x$. For macroparticules distributions extracted from PIC simulations, the weights $w_i$ of each macroparticle have to be taken into account :
\[\left\langle x^2 \right\rangle  = \frac{\sum\nolimits_i \left( w_i(x_i - \bar x)^2 \right) }{\sum\nolimits_i w_i }\] and
\[\left\langle x\:p_x \right\rangle  = \frac{\sum\nolimits_i \left( w_i(x_i - \bar x)(p_{x_i} - \bar p_x) \right) }{\sum\nolimits_i w_i }\]
where $\bar x$ and $\bar p_x$ are respectively the weighted-average of $x$ and $p_x$ :
\[\bar x = \sum\nolimits_i  \left( w_i x_i \right) /\sum\nolimits_i  w_i \quad {\rm and} \quad \bar p_x =\sum\nolimits_i  \left( w_i p_{x_i} \right) /\sum\nolimits_i  w_i.\]
The definitions are similar for the emittance calculation in the $(y,p_y)$ plan.


%

\section*{Acknowledgments}

We gratefully acknowledge the support of the LULI pico 2000 staff during the experimental run. J.J.S. gratefully acknowledges fruitful discussions with L. Gremillet. This work was performed through funding from the French National Agency for Research (ANR) and the competitiveness cluster Alpha - Route des Lasers, project number TERRE ANR-2011-BS04-014. The authors also acknowledge support from the COST Action MP1208 "Developing the physics and the scientific community for Inertial Fusion" through three STSM visit grants. The research was carried out within the framework of the "Investments for the future" program IdEx Bordeaux LAPHIA (ANR-10-IDEX-03-02) and of the EUROfusion Consortium and has received funding from the European Union's Horizon 2020 research and innovation program, grant 633053. The views and opinions expressed herein do not necessarily reflect those of the European Commission. The Japanese collaborators were supported the Japanese Ministry of Education, Science, Sports, and Culture through Grants-in-Aid for Young Scientists (Grants No. 24684044), Grants-in-Aid for Fellows by JSPS (Grant No. 14J06592), and the program for promoting the enhancement of research universities.
Simulation work has been partially supported by the Spanish Ministry of Economy and Competitiveness (grant No. ENE2014-54960-R) and used HPC resources and technical assistance from CeSViMa Centre of the UPM.

\section*{Author Contributions}
J.J.S. designed and executed the experiment as principal investigator, with the help from M.B.-G., C.B., P.F.-C., S.F., L.G., J.E.C., R.C., J.-L.D., M.E., S.K., J.-R.M., A.M., S.S., J.S. and Z.Z. and the engineering support from R.B., M.C., S.D., E.L. and F.S.; data analysis was performed by M.B-G., with contributions from P.F-C., L.G., M.E., S.K., S.S. and Z.Z. and under the supervision of J.S.S., D.B. and S.F.; B-field diffusion modelling was performed by J.J.H., who also developed the PIC-hybrid code; the CTR post-processor was developed by C.B.; PIC-hybrid simulations were run and analysed by M.B.-G. with supervision of J.J.S, C.B. and J.J.H.; targets were manufactured by G.S.; G.G., S.H., Ph.N., M.R., V.T.T. and N.W. contributed to the discussion of the results; M.B.-G. and J.J.S. led the manuscript writing; all the figures were prepared by M.B.-G.    




\bibliographystyle{unsrt}

\end{document}